\documentclass[prb,twocolumn]{revtex4}
\usepackage{graphicx}
\usepackage{bm}
\usepackage{amsmath}
\usepackage{indentfirst}
\usepackage{times}
\usepackage{hhline}

\newcommand{\ph}[1]{\phantom{#1}}

\begin{document}

\title{Polarization enhancement in two- and three-component
ferroelectric superlattices}

\author{S. M. Nakhmanson}
\author{K. M. Rabe}
\author{David Vanderbilt}

\affiliation{Department of Physics and Astronomy, Rutgers
University, Piscataway, NJ 08854-8019}

\date{\today}

\begin{abstract}
Composition-dependent structural and polar properties of epitaxial
short-period CaTiO$_3$/SrTiO$_3$/BaTiO$_3$ superlattices grown on
a SrTiO$_3$ substrate are investigated with first-principles
density-functional theory computational techniques. Polarization
enhancement with respect to bulk tetragonal BaTiO$_3$ is found for
two- and three-component superlattices with a BaTiO$_3$ concentration
of more than 30\%. Individual BaTiO$_3$ layer thickness is
identified as an important factor governing the polarization
improvement.  In addition, the degree of inversion-symmetry breaking in
three-component superlattices can be controlled by varying the
thicknesses of the component layers. The flexibility allowed within
this large family of structures makes them highly suitable for
various applications in modern nano-electro-mechanical devices.
\end{abstract}

\maketitle

With modern state-of-the-art epitaxial engineering techniques,
complex perovskite oxide superlattices with three different consituent 
layers can be grown to realize a wide range of designer materials for 
a variety of applications.~\cite{yamada2002,eckstein2003,lee2005}
Recently, it was shown that hundreds of atomically thin individual layers of
CaTiO$_3$ (CT), SrTiO$_3$ (ST) and BaTiO$_3$ (BT) could be grown on
a pe\-rov\-skite ST substrate, yielding superlattices with
compositionally abrupt interfaces, atomically smooth surfaces and
excellent polar properties.~\cite{eckstein2003,lee2005} Since the 
relaxed lattice constants of CT and BT are 0.07 \AA\ smaller and 0.11
\AA\ larger than that of ST (3.905 \AA), respectively, epitaxial
strain in such structures results in substantial polarization
enhancement.~\cite{neaton2003} In addition, the inversion symmetry
breaking present in three-component 
superlattices~\cite{eckstein2003,lee2005,sai2000}
leads to even greater flexibility in fine-tuning the ferroelectric
properties of these materials.

In this letter, in the spirit of Ref.~[\onlinecite{neaton2003}], we
investigate structural and polarization-related properties of
two-component (or ``bicolor'') and three-component (``tricolor'')
(ST)$_l$(BT)$_m$(CT)$_n$ ferroelectric superlattices, where $l,n$ =
0,1,2 and $m = 0,\ldots,4,$ epitaxially matched to a cubic ST
substrate.  We demonstrate that a number of such structures have
enhanced spontaneous polarization with respect to bulk tetragonal
BT. We also show that, in agreement with the findings of
Ref.~[\onlinecite{neaton2003}], all superlattice layers, including
the naturally-paraelectric ST ones, are strongly
polarized, resulting in a smooth polarization profile along the
$[001]$ (epitaxial growth) direction in the structure.

All the calculations presented here were performed under periodic
boundary conditions equivalent to the presence of short-circuited
top and bottom ``electrode'' layers. A plane-wave based DFT-LDA
method~\cite{pwscf} with ultrasoft pseudopotentials~\cite{USPP} was
utilized for the ionic relaxation of the \hbox{$1 \times 1 \times
(l+m+n)$} supercells.  30 Ry wavefunction and 270 Ry electronic
density plane-wave cutoffs were used in all the calculations.
During the relaxations, the in-plane lattice constant $a$ of the
tetragonal cell was constrained to the theoretical lattice constant
of cubic ST (3.858 \AA\ in this investigation) and the out-of-plane
lattice constant $c$ was allowed to vary. The symmetry in all the
calculations was restricted to space group P4mm (point group
C$_{4v}$), i.e., the ions could move only in the [001] direction. The
system was considered to be at equilibrium when the Hellman-Feynman
forces on the ions were less than \hbox{0.5 $\times$ 10$^{-3}$ Ry/bohr}
and the $\sigma_{33}$ component of the stress-tensor was smaller
than 0.5 KBar. We used a \hbox{$6 \times 6 \times N$}
Monkhorst-Pack (MP) mesh~\cite{kpt} for all the Brillouin-zone
integrations, where $N = 6/(l+m+n)$ for $l+m+n \leq 3$, $N = 2$ for
$l+m+n = 4$ and $N = 1$ for $l+m+n \geq 4$. We employed the
Berry-phase method of the modern polarization theory~\cite{mpt} to
compute the total polarization of each superlattice. The \hbox{$6
\times 6 \times 2N$} MP mesh used in the polarization calculations
produced well-converged results.

Calculated structural parameters for the bicolor and tricolor
(ST)$_l$(BT)$_m$(CT)$_n$ superlattices grown on ST, as
well as for strained bulk CT and BT and unstrained bulk tetragonal
BT, are presented in Table~\ref{data}. We find that bulk BT, with
a relaxed lattice constant 2\% larger than that of cubic ST, expands
by 5.5\% in the [001] direction when pseudomorphically constrained
to the ST substrate. Analogously, CT, with relaxed lattice constant
1.2\% smaller than that of the substrate, contracts by about 1\%.
These results are in excellent agreement with calculations of
Di\'eguez \textit{et al.}~\cite{dieguez2005} The $c/a$ ratios for
the bicolor and tricolor superlattices show similar trends
corresponding to BT layers expanding and CT layers slightly
contracting. The tricolors have two sets of tabulated values
because the lack of inversion symmetry in these systems makes
ferroelectric displacements along [001] and [00$\bar{1}$]
inequivalent. It is worth pointing out that the $c/a$ values in the
tricolor systems are slightly different for the [001] and [00$\bar{1}$]
displacements, reflecting the presence of a polarization-strain
coupling in the films. The experimental values for $c/a$
ratios,~\cite{ho_nyung} shown in parentheses for some structures, are
in excellent agreement with the results of our calculations, with
the only minor exceptions being for (BT)$_1$(CT)$_1$ and tetragonal BT.
Experimental parameters for the latter were obtained for a
thin-film sample grown under the same conditions as the
superlattice samples. The most likely source of discrepancy
between the theoretical and experimental parameters in both of
these cases could be the structural defects present in the
experimentally grown thin films.

\begingroup
\begin{table}[t!]
\tabcolsep=0.16cm
\renewcommand{\arraystretch}{0.9}

\caption{
Structural parameters and polarization in bulk CT and BT, and
in (ST)$_l$(BT)$_m$(CT)$_n$ superlattices, pseudomorphically grown on
ST (with in-plane lattice constant $a$ = 3.858 \AA).
$P_{[001]}^{\mathrm{BP}}$ was computed with the Berry-phase method.
$|P|/P_{\mathit{t}\mbox{-}\mathrm{BT}}$ is the polarization
enhancement factor with respect to the polarization in bulk
tetragonal BT. Experimental values, included for comparison, are
shown in parentheses where available. The positive direction of the
$[001]$ axis is from the ST to the neighboring CT layer.
}
\label{data}
\begin{ruledtabular}
%\begin{center}
\begin{tabular}{lccc}

  System    &    $c/a$   & $P_{[001]}^{\mathrm{BP}}$ (C/m$^2)$ &
  $|P|/P_{\mathit{t}\mbox{-}\mathrm{BT}}$ \\
 		       		       
\hline		       		       
		       		       
  Strained bulk:   \\

  CT                      &  0.9897     & \ph{-}0.434       & \\

  BT                      &  1.0548     & \ph{-}0.368       & \\

\hline

  Bicolors:   \\

  (ST)$_1$(CT)$_1$               &  1.976 (1.97)   &  \ph{-}0.026 (0.000)   & 0.11 (0.0) \\

  (ST)$_1$(BT)$_1$               &  2.042 (2.07)   &  \ph{-}0.231 (0.059)   & 0.95 (0.54) \\

  (BT)$_1$(CT)$_1$               &  2.019 (2.08)   &  \ph{-}0.231 (0.085)   & 0.95 (0.78) \\

  (ST)$_2$(CT)$_2$               &  3.960          &  \ph{-}0.168           & 0.69 \\

  (ST)$_2$(BT)$_2$               &  4.088          &  \ph{-}0.245           & 1.01 \\

  (BT)$_2$(CT)$_2$               &  4.059          &  \ph{-}0.306           & 1.26 \\

\hline

  Tricolors:   \\					    				    

  (ST)$_1$(BT)$_1$(CT)$_1$      &  3.018 \raisebox{-5pt}[1pt][1pt]{(3.02)} 
                                & -0.200 \raisebox{-5pt}[1pt][1pt]{(0.030)}   
                                &  0.82  \raisebox{-5pt}[1pt][1pt]{(0.28)}  \\

                                &  3.015 \ph{(3.02)}              
                                & \ph{-}0.171  \ph{(0.030)}           
                                &  0.70 \ph{(0.28)} \\

  (ST)$_2$(BT)$_2$(CT)$_2$      &  6.049 \raisebox{-5pt}[1pt][1pt]{(6.05)}    
                                & -0.239 \raisebox{-5pt}[1pt][1pt]{(0.069)}   
                                &  0.98  \raisebox{-5pt}[1pt][1pt]{(0.63)} \\

                                &  6.051 \ph{(6.05)}
                                & \ph{-}0.242  \ph{(0.069)}
                                &  1.00 \ph{(0.63)} \\

  (ST)$_2$(BT)$_4$(CT)$_2$      &  8.163 \raisebox{-5pt}[1pt][1pt]{(8.21)}   
                                & -0.295 \raisebox{-5pt}[1pt][1pt]{(0.133)}   
                                &  1.21  \raisebox{-5pt}[1pt][1pt]{(1.22)} \\

                                &  8.165 \ph{(8.21)}
                                & \ph{-}0.298  \ph{(0.133)}
                                &  1.23 \ph{(1.22)} \\

  (ST)$_3$(BT)$_3$(CT)$_3$      &  9.083           
                                & -0.260            
                                &  1.07 \\

                                &  9.077           
                                & \ph{-}0.244            
                                &  1.00 \\

\hline

  Bulk tetragonal BT            &  1.009 (1.026)   & \ph{-}0.243 (0.109)    &  \\

\end{tabular}
%\end{center}
\end{ruledtabular}
\end{table}
\endgroup

Turning to the polar properties of the superlattices, in columns 3
and 4 of Table~\ref{data} we assemble spontaneous polarizations
$P_{[001]}^{\mathrm{BP}}$, computed with the Berry-phase method, as
well as polarization enhancement factors with respect to bulk
tetragonal BT $|P|/P_{\mathit{t}\mbox{-}\mathrm{BT}}$ for all the
superlattices and bulk materials studied in this investigation.
Starting at the top of the Table, we notice that our calculations
provide a very high polarization for strained bulk CT. With the
aforementioned symmetry restrictions in place, this is the
polarization CT \textit{would have} if the zone boundary
distortions that are present in its actual ground-state crystal
structure were suppressed. The DFT-based model of Di\'eguez
\textit{et al}~\cite{dieguez2005} produces a similar result for the
strained CT.  However, this peculiar property does not
automatically translate into highly polar CT-containing superlattices.
While CT and BT are both ferroelectric in the (constrained) bulk,
they behave quite differently in the superlattice geometries.
BT is highly polar in monoatomic layer systems like
(ST)$_1$(BT)$_1$ or (ST)$_1$(BT)$_1$(CT)$_1$, and becomes even more
polar with growing BT-layer thickness $m$. This can be seen from
the data for the ``constant-concentration'' (ST)$_m$(BT)$_m$ and
(BT)$_m$(CT)$_m$ bicolor systems, and substantiated by the
(ST)$_m$(BT)$_m$(CT)$_m$ tricolor series as well as the BT-rich
(ST)$_2$(BT)$_4$(CT)$_2$ system. CT, at variance, is
practically paraelectric in the monoatomic layer systems, with
(ST)$_1$(CT)$_1$ exhibiting low polarization and (BT)$_1$(CT)$_1$
showing the same polarization as (ST)$_1$(BT)$_1$. However, as the
thickness of the CT layer ($n$) increases, so does its polarization:
the (ST)$_2$(CT)$_2$ system is substantially more polar than
(ST)$_1$(CT)$_1$, and (BT)$_2$(CT)$_2$ is more polar than
(ST)$_2$(BT)$_2$.  In (ST)$_3$(BT)$_3$(CT)$_3$, the ferroelectric
displacements in the CT layer are larger than those in the BT
layer. These results suggest that CT and BT cells tend to remain
more polar when assembled into thick layers, as opposed to being
intermixed with each other and naturally paraelectric
ST.~\cite{serge} However, in experimentally studied systems,
epitaxial growth restrictions place a limit on the values of $n$
and $m$.~\cite{schlom2004} Above a certain thickness, CT or BT layers
relax to their native in-plane lattice constants and the
strain-induced polarization enhancement is lost. For example, in
the (ST)$_2$(BT)$_m$(CT)$_2$ superlattices, the polarization starts to
decrease at $m \simeq 7,$~\cite{lee2005,ho_nyung} so that a
delicate balance between concentration and layer thickness has to
be maintained to produce superlattices with the strongest
polarization enhancement. Finally, there is an additional degree of
tunability of polar properties in the tricolor superlattice family. As
shown in the bottom part of the Table, the inequivalence of the
[001] and [00$\bar{1}$] ferroelectric displacement directions
results in two distinct values of polarization in these
structures.

Experimental values for the spontaneous polarizations in some
ferroelectric superlattices, obtained by Lee \textit{et
al},~\cite{lee2005,ho_nyung} together with the associated
polarization enhancement factors are shown for comparison in
Table~\ref{data}. The former are consistently smaller than the
polarizations obtained in the calculations. However, during the
experimental sequences of poling and polarization reversal,
structural defects and incomplete switching of ferroelectric
domains usually do lead to substantially reduced values of remanent
polarization. Our calculations, on the other hand, correspond to an
``ideal case'' (no defects and perfect switching) and thus provide
upper-bound polarization estimates.  Nevertheless, if we compare
enhancement factors instead of actual polarizations, we see certain
similar trends
in the experimental and theoretical data. For example, experimental
results for the tricolor systems show the same sequence of
polarization enhancement (from (ST)$_1$(BT)$_1$(CT)$_1$ to
(ST)$_2$(BT)$_4$(CT)$_2$) as the calculated one. Also the agreement
between theoretical and experimental polarization-enhancement factors 
in tricolor systems improves with increasing polarization of the system. 
The predicted difference in polarizations induced by inequivalent
ionic displacements in the tricolor superlattices has not yet been
observed in the samples of Lee \textit{et al} due to other sources
of asymmetry in the experimental geometries.\cite{lee2005,ho_nyung}

Since the intricate relation between pseudomorphic strain and
chemical composition of the layers is the key to the polarization
enhancement in the superlattices, we investigated the
layer-by-layer polarization behavior in these structures.
Density-functional perturbation theory~\cite{dfpt} was used to
obtain Born effective charges for ions in relaxed supercells, and
these were then used to
decompose an ``aggregate'' supercell polarization into
contributions from individual primitive cells comprising each
superlattice. The polarization contribution $P_{\lambda}$ from
cell $\lambda$ was estimated using the linearized
approximation
\begin{equation}
P_{\lambda} \simeq \sum_i  \frac{\partial P}{\partial u_{\lambda i}^{(0)}} 
( u_{\lambda i} - u_{\lambda i}^{(0)} )
= \frac{1}{V_{\lambda}} \sum_i Z^*_{\lambda i} \Delta u_{\lambda i},
\label{polar_eq}
\end{equation}
where $Z^*_{\lambda i}$ and $\Delta u_{\lambda i}$ are the effective
charge and displacement of ion $i$ in cell $\lambda$, and
$V_{\lambda}$ is the volume of the cell. All polarizations and
ionic displacements are constrained to the [001] direction, and
a superscript zero refers to a non-polar structure with
ferroelectric displacement removed by ``unbuckling'' the AO and BO
planes and moving the ``middle'' planes back to the center of each
primitive cell. Equation~(\ref{polar_eq}) can be used with either A
or B cation-centered cells. In each case we average out
contributions of the oxygens in the planes bounding elementary
cells, while cations at the corners set the volume of each cell.
The total supercell polarization $P_{[001]}^{\mathrm{BEC}}$ can be
recovered from Eq.~(\ref{polar_eq}) by extending the summation
to all the ions in the system. Total polarizations obtained by such
summations are in good agreement with those computed by the
Berry-phase method.

\begin{figure}[t!]
\begin{center}
\includegraphics[scale=.53]{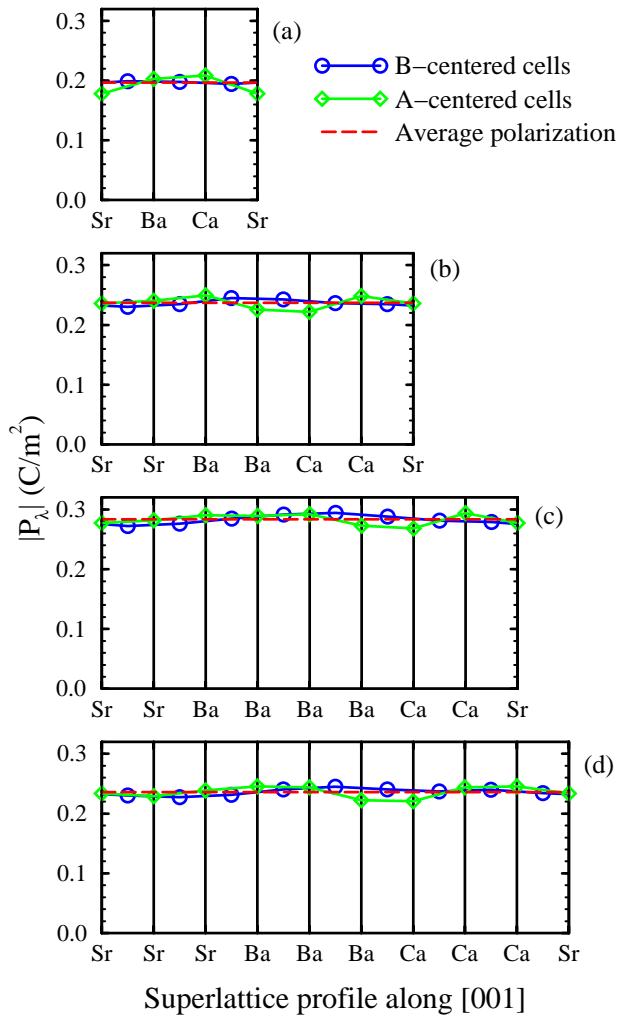}
\caption{Examples of local polarization profiles of the tricolor
(ST)$_l$(BT)$_m$(CT)$_n$ superlattices: (a) $l,m,n=1$; (b)
$l,m,n=2$; (c) $l,n=2$, $m=4$; (d) $l,m,n=3$.  Dashed lines show
the absolute value of total average polarization in the supercells
($P_{[001]}^{\mathrm{BEC}}$).
}
\label{fig1}
\end{center}
\end{figure}

In Fig.~\ref{fig1} we present the layer-by-layer polarization
profiles of the tricolor superlattices from Table~\ref{data}. Both
A and B cation-centered cell profiles are shown for each
superlattice, but the differences between the two are small. Both
types of analysis show that all of the layers in the superlattice
structures (including the naturally paraelectric ST ones)
are strongly polarized, and that the local polarization is almost
uniform throughout the superlattice.  As already pointed out in
Ref.~[\onlinecite{neaton2003}], the near constancy of polarization
throughout the superlattice results from the minimization of the
energy terms associated with the polarization charge build-up
($\mathbf{\nabla\cdot P}$) at the interfaces (i.e., it is energetically
unfavorable for the system to have a large divergence of the polarization).

In summary, we have used first-principles methods to study structural
properties and polarization enhancement in short-period
perovskite-type thin films containing individual layers of CT, ST
and BT. We find that the presence of highly polar CT and BT layers
induces substantial polarization in these superlattices. The actual
degree of polarization enhancement strongly depends on an interplay
between the concentration of ferroelectric components in the
superlattice and the pseudomorphic strain it can sustain. Maximum
polarization enhancement is achieved by the highest concentration
of CT or BT assembled into the thickest possible layers that are
still consistent with a state of full epitaxial strain. Our
calculations also show that the tricolor superlattices that lack
a center of inversion have two distinct values of polarization
depending on the direction of the ferroelectric displacement, which
can be exploited for fine-tuning of the polar properties. We hope
that this investigation, by providing additional insight
into the nature of complex polar perovskite materials, will lead to
better understanding and control of their properties in the quest
for more efficient and environmentally friendly
nano-electro-mechanical devices.

The authors thank H. N. Lee for sharing his data and many valuable
discussions. This work was supported by the Center for
Piezoelectrics by Design (CPD) under ONR Grant N00014-01-1-0365.


\begin{thebibliography}{99}
\parskip = 0pt

%\bibitem{schlom99} J. C. Jiang, X. Q. Pan, W. Tian, C. D. Theis and
%D. G. Schlom, Appl. Phys. Lett. \textbf{74}, 2851 (1999).

\bibitem{yamada2002} H. Yamada, M. Kawasaki, Y. Ogawa, and Y. Tokura,
Appl. Phys. Lett. \textbf{81}, 4793 (2002).

\bibitem{eckstein2003} M. P. Warusawithana, E. V. Colla, J.N. Eckstein,
and M. B. Weissman, Phys. Rev. Lett. \textbf{90}, 036802 (2003).

\bibitem{lee2005} H. N. Lee, H. M. Christen, M. F. Chisholm, C. M.
Rouleau and D. H. Lowndes, Nature \textbf{433}, 395 (2005).

\bibitem{neaton2003} J. B. Neaton and K. M. Rabe, Appl. Phys. Lett.
\textbf{82}, 1586 (2003).

\bibitem{sai2000} N. Sai, B. Meyer and D. Vanderbilt, Phys. Rev.
Lett. \textbf{84}, 5636 (2000).

\bibitem{pwscf} We used PWscf code (available from
http://www.pwscf.org) for the calculations presented here.

\bibitem{USPP}  D. Vanderbilt, Phys. Rev. B \textbf{41}, 7892
(1990).

\bibitem{kpt} H. J. Monkhorst and J. D. Pack, Phys. Rev. B
\textbf{13}, 5188 (1976).

\bibitem{mpt} R. D. King-Smith and D. Vanderbilt, Phys. Rev. B
\textbf{47}, 1651 (1993); R. Resta, Rev. Mod. Phys. \textbf{66},
899 (1994).

\bibitem{dieguez2005} O. Di\'eguez, K. M. Rabe and D. Vanderbilt,
to be published.

\bibitem{ho_nyung} H. N. Lee, private communication.

\bibitem{serge} This is conceptually very similar to polarization
behavior in polymer ferroelectrics. See S. M. Nakhmanson. M.
Buongiorno-Nardelli and J. Bernholc, to be published.

\bibitem{schlom2004} K. J. Choi, M. Biegalski, Y. L. Li, A. Sharan,
J. Schubert, R. Uecker. P. Reiche, Y. B. Chen, X. Q. Pan, V.
Gopalan, L.-Q. Chen, D. G. Schlom, C. B. Eom, Science \textbf{306},
1005 (2004);

\bibitem{dfpt} S. Baroni, S. de Gironcoli, A. Dal Corso and P.
Giannozzi, Rev. Mod. Phys.  \textbf{73}, 515 (2001).

\end{thebibliography}
\end{document}